\documentclass[12pt,preprint]{aastex}
\usepackage[english]{babel}
\usepackage{subfigure}
\usepackage{natbib}

\shorttitle{Tidal Stripping of Globular Cluster}
\shortauthors{Coenda et al.}

\newcommand{\kpc}{{\rm kpc}}
\newcommand{\Mpc}{{\rm Mpc}}
\newcommand{\kms}{{\rm km\,s^{-1}}}
\newcommand{\keV}{{\rm keV}}

\def\arcmin{\rm arcmin}
\def\arcsec{\rm arcsec}
\newcommand{\Msun}{{\mathcal{M}_{\odot}}}

\def\kms {\rm{km~s^{-1}}}

\begin{document}

\title{Tidal Stripping of Globular Clusters in the Virgo Cluster}

\author{Valeria Coenda\altaffilmark{1}, Hern\'an Muriel\altaffilmark{1}}
\affil{Instituto de Investigaciones en Astronom\'\i a Te\'orica y Experimental
IATE, Observatorio Astron\'omico OAC, Laprida 854, X5000BGR, C\'ordoba, Argentina.}
\email{vcoenda@oac.uncor.edu, hernan@oac.uncor.edu}
\and
\author{Carlos Donzelli\altaffilmark{1,2}}
\affil{Space Telescope Science Institute, 3700 San Martin Drive, Baltimore, MD 21218, USA. IATE, OAC, Argentina.}
\email{donzelli@stsci.edu}

\altaffiltext{1}
{ Consejo Nacional de Investigaciones Cient\'\i ficas y T\'ecnicas (CONICET),
Avenida Rivadavia 1917, C1033AAJ, Buenos Aires, Argentina.}
\altaffiltext{2}
{AURA US Gemini Fellow}

\begin{abstract}
With the aim of finding evidence of tidal stripping of globular clusters (GCs)
we analysed a sample of 13 elliptical galaxies taken from the ACS
Virgo Cluster Survey (VCS). These galaxies belong to the main concentration of
the Virgo cluster (VC) and present absolute magnitudes $-18.5<M_z<-22.5$.
We used the public GC catalog of Jord\'an et al.
(2008) and separated the GC population into metal
poor (blue) and metal rich (red) according to their integrated colors.
The galaxy properties were taken from \citet{Peng:2008}.
We found that: 

1) The specific frequencies ($S_N$) of total and blue 
GC populations increase as a function of the projected galaxy distances $r_p$ to M87. 
A similar result is observed when 3-dimensional distances $r_{3D}$ are used. The same behaviours 
are found if the analysis are made using the number of GCs per $10^9\Msun$ ($T$). 
No correlations between $S_N$ or $T$ and $r_p$ or $r_{3D}$ is observed for the red GC  
population. The correlations for the blue GCs (typically more extended) and the lack 
of correlations for the red GCs (more concentrated) with the clustocentric distance  
of the host galaxy are interpreted as evidence of GCs stripping due to tidal forces. 

2) No correlation is found between the slope of GC density profiles 
of host galaxies and the galaxy distance to M87 (Virgo central galaxy). 
The lack of such a correlation is interpreted in terms of a shrinkage of the GC distribution after the stripping of GCs in the outermost region of galaxies. 

3) We also computed the local density of GCs ($\rho_{out}$) located
further than $6.2 \,\kpc$ from the
galaxy center for nine galaxies of our sample. We find that the GC
population around most of these
galaxies is mainly composed of blue GCs. The two highest values of
$\rho_{out}$ are found in the
core of the VC (up to $100\,\kpc$) and correspond to the two lowest
values of $S_N$.

Our results suggest that the number and the fraction of blue and red GCs
observed in elliptical galaxies located near the centers of massive clusters,
could be significantly different from the underlying GC population.
These differences could be explained by tidal stripping effects that
occur as galaxies approach the centers of clusters.
\end{abstract}



\section{Introduction}\label{Intro}

It has been demonstrated that tidal interactions can strongly affect
the evolution of galaxies in clusters.  Rapid gravitational encounters
(galaxy harassment) as well as the global gravitational field of the
cluster itself can dramatically change galaxy properties. Some of the
properties that can be affected by tidal forces are those related to
the population of globular clusters (GCs). Muzzio and collaborators in
the 80s (\citealt{Muzzio:1986}, \citealt{Muzzio:1987}), and more recently Bekki et
al. (2003, hereafter B2003), \nocite{Bekki:2003} used collisionless dynamical simulations
to demonstrate that galaxies in clusters can lose a significant fraction of their GC population as a consequence of tidal stripping.
Moreover, since the outermost regions of a galaxy are more sensitive
to tidal effects, B2003 found that the number density profile of the
GC population becomes steeper after the stripping. In addition,
given that tidal encounters as well as the global effect of the cluster
potential increase towards the inner region of the cluster, a positive
correlation between the slope of the density profile of the GC population and the
clustocentric distance of the galaxies is expected to be observed. 

If GCs are stripped from galaxies, a fraction of these GC should populate
the intracluster medium. Bekki \& Yahagi (2006, hereafter BK2006) \nocite{B&Y:2006} 
investigated the spatial distribution of intracluster GCs (ICGC) using high-resolution 
cosmological simulations, and found that ICGC population contribute 20-40\% to the total
GC population in high mass clusters, making the case for an inhomogeneous, asymmetric
and elongated distribution.

It is well known that color distributions of GCs are typically bimodal
\citep{Peng:2006}, indicating two sub-populations of GCs.
Spectroscopic studies have shown that color bimodality is mainly due
to differences in the metallicity of the two sub-populations: the
metal-poor GCs are blue, while the metal-rich GC are red. Blue GCs
present a shallower and more extended radial density profile than red GCs,
and consequently have a higher probability to be stripped from the
parent galaxy. In the tidal stripping scenario, the cluster core and,
therefore, the central galaxy play an important role. For example,
\citet{Forte:1982} found that the properties of the GC population in
M87 (Virgo central galaxy) are consistent with the accretion of GCs from less massive
galaxies. \citet{Cote:1998} explored the possibility that the
metal-poor population of giant elliptical galaxies arises from the
capture of GC from other galaxies. Analysing the two brightest
galaxies in Virgo, these authors concluded that it is possible to
explain the observed properties of the blue GC population in M49 and M87
without invoking new GC formation through mergers or multiple bursts.

\citet{Forbes:1997} found evidence of tidal stripping in Fornax.
These authors analyzed the specific frequency of GCs in four galaxies
located in central cluster regions and found a marginal dependence of
$S_N$ (the number of GCs scaled to the galaxy luminosity) on the clustocentric 
galaxy distance $r_c$. Nevertheless, the $r_c$ range considered is relatively small ($r_c < 200 \kpc$). The same cluster was also analysed by \citet{Hilker:1999}. These authors
suggested that an important fraction of the GC population of the cD
galaxy NGC 1399 could be explained by GCs stripped from the central
giant galaxies making those central galaxies to present low specific frecuencies.
In order to quantify the importance of the tidal stripping effect, it is
fundamental to study new clusters of galaxies and also to extend the 
limited range of clustocentric distances currently available. 
Recently, Peng et al. (2008, hereafter P2008) studied the formation efficiencies 
of GCs in the Virgo Cluster Catalog (VCC) sample. They found that both, high and low luminosity early
type galaxies, have large $S_N$ values, while those galaxies with
intermediate luminosities ($-20<M_B<-17$) present small and
relatively constant $S_N$. They also found that $S_N$ depends on the
environment, i.e. 1) galaxies within a projected radius of 1 Mpc from the
cluster center tend to have higher GC fractions, and 2) galaxies within $r_p\simeq 40\,\kpc$
of M87 (and M49) have few or no GCs.
The first result is interpreted in terms of the formation time, where GCs formation in
central dwarf galaxies is biased because their stars form earlier; whereas the second result 
considers that GCs are tidally stripped by their giant neighbors.

There are several processes that can produce tidal stripping of GCs.
The main candidates are the pure gravitational forces due to the cluster potential well
(see B2003), and the galaxy-galaxy interaction (galaxy harassment). 
Using numerical simulations, \citet{Moore:1996} demonstrated that in high mass galaxy cluster, 
this effect can indeed produce the stripping of an important part of the stellar component 
of galaxies. Based on the analysis of 228 elliptical galaxies in clusters, \citet{Cypriano:2006} found that galaxies in the inner region of clusters
are 5\% smaller than those in the outer regions. This result is interpreted as evidence 
of the stripping of stars from elliptical galaxies in the central regions of clusters.

A precise knowledge of the number of GCs that galaxies could lose due to tidal stripping is
particularly interesting for several reasons:
i) If the models of GCs formation are going to be tested using observational data, GC population
should be corrected by losses due to tidal stripping.
ii) Some models for the formation and evolution of cD galaxies assume that
a fraction of their GC population comes from tidal stripping of other
cluster galaxy members (\citealt{Forte:1982}, \citealt{Hilker:1999}). The confirmation that tidal 
stripping is an efficient process will support this theory.
iii) The GC population extends to large radii from the center of a galaxy. Therefore, GCs are ideal to study the outskirts of galaxies. If we can observe that 
GCs are being stripped by tidal interactions, this effect can be used to test a similar 
effect in other components of a galaxy, such as the dark matter halo.

In this work, we study the GC population of a sample of 13 member
galaxies of the Virgo Cluster (VC). The ACS Virgo Cluster Survey (ACS VCS) is described in C{\^o}t{\'e} et al. (2004) (hereafter C2004)\nocite{Cote:2004}. 
There are many relevant papers produced by the ACS team: Color distributions of the GCs are presented 
in \citet{Peng:2006}; GC size distributions are analysed in \citet{Jordan:2005}; 3-dimensional 
distances for 84 galaxies in the VC are presented in \citet{Mei:2007}; surface brightness profiles, 
total magnitudes and colors of the sample galaxies are described in \citet{Ferrarese:2006};
specific frequency, GC number, $T$ parameter, and other galaxy parameters are presented 
in P2008, while the GC catalogue used in the present work is described in \citet{Jordan:2008}.
The aim of this paper is to detect whether there exist observational evidence of the tidal
stripping effect on the GC population due to galaxy-galaxy
interactions and/or by the cluster potential well. The paper is
organized as follows: sample selection is described in Section
\ref{Sample}, we discuss the GC selection in Section \ref{gc}, while in Section \ref{radial}, we derive and analyze our results; the final conclusions are 
summarized in Section \ref{conclusion}.

\section{The Sample}\label{Sample}

\subsection{Galaxy Sample}\label{galaxies}

The Virgo Cluster (VC) is a well studied cluster at a distance of 16.5$\Mpc$ (\citealt{Tonry:2001}, 
\citealt{Mei:2007}). It has a velocity dispersion of $776\pm 21$ $\kms$
\citep{Girardi:1993}, and an intracluster gas temperature of 2.4
$\keV$ \citep{David:1993}.  Therefore, this cluster is an excellent
candidate to study tidal stripping effects on the GC
population. The ACS Virgo Cluster Survey is a dedicated optical and IR imaging survey of 100 Virgo's
early-type galaxies, which made use of the {\it Advanced Camera for Surveys} (ACS) 
on-board the {\it Hubble Space Telescope} (HST).
These observations provide a unique opportunity to study GCs in cluster galaxies, as done by 
the ACS VCS team. They also provide an insight into the GC population in the intracluster medium (ICM).
The complete survey is described in \citet{Cote:2004} and data 
reduction techniques in \citet{Jordan:2004}. Briefly, the galaxies were observed
through the F475W and F850LP filters that correspond to the Sloan $g$
and $z$ filters, respectively. The sample includes galaxies 
with $B_T<16$ and they were required to have morphological types of E, S0, dE, dE,N, dS0 or dS0,N.

The number of GCs depends on the luminosity of the host galaxy. Moreover, the specific frequency $S_N$
depends on the galaxy magnitude (\citealt{Lotz:2004}, \citealt{Miller:2007}, P2008).
Since the ACS VCS sample spans over a wide luminosity range, and we want to avoid any luminosity 
bias in our analysis, we selected those galaxies with $-22.5<M_z<-18.5$. Over this luminosity range, the 
specific frequency $S_N$ and the $T$ parameter (defined in Sect. \ref{radial}) do not depend on 
the galaxy magnitude (P2008). On the other hand, since the GC population could depend on the host 
galaxy morphology, we have selected only elliptical galaxies. This type of galaxies shows (on average)
the highest GC specific frequency for a given luminosity. In addition, we 
have only considered those elliptical 
galaxies that belong to the main cluster structure around M87 \citep{Mei:2007}. The last criterion 
excludes VCC1178 and VCC1025 since they belong to the cluster sub-structure associated to M49. 
The sample comprises 13 elliptical galaxies located around M87. They span a range of galaxy projected distance to M87 between $0.04$ and 
$1.65\,\Mpc$, while their total absolute magnitudes satisfy that $-22.5<M_z<-18.5$. Several galaxy properties,as the specific frecuency $S_N$, $T$ parameter and other properties, were taken from P2008. We have also used the 3-dimensional distances $r_{3D}$ obtained by \citet{Mei:2007}. For our sample we obtain clustercentric distances $0.23<r_{3D}<2.0$. It should be taken into account that these distances have a line-of-sight 
uncertainties of $\sim$0.6 Mpc. \citet{Mei:2007} adopted the standard $\Lambda$CDM cosmology with $\Omega_m h^2=0.127_{-0.013}^{+0.007}$ and $h = 0.73 \pm 0.03$. Details of the 13 selected galaxies are quoted in Table \ref{tb:data}.

\subsection{Globular Cluster Selection}\label{gc}

We specifically used the catalog of GC published by \citet{Jordan:2008}.
This catalog contains 12763 bona-fide GC candidates which have the probability
$p_{GC} > 0.5$ of being a GC according to their selection procedure.
Basically, the selection procedure involves three main steps. The first one is described
in \citet{Jordan:2004} and contemplates the modeling and removal of the galaxy surface
brightness distribution and subsequent object detection performed with SExtractor
\citep{Sextractor:1996}. In the second step all the GC candidates are run through the
KINGPHOT code that fits PSF-convolved King models \citep{King:1966} to their surface brightness
profiles. Details about this code can be found in \citet{Jordan:2005}.
Evaluated parameters for each GC candidate are: magnitude $z$ (total and in a $0.2\,\arcsec$ diaphragm),
celestial coordinates $\alpha$ and $\delta$, half-light radius ($r_h$) and concentration defined
as the logarithm of the core and tidal radii ratio. However, after the initial selection of GC
candidates mentioned above there are still residual contaminants such as foreground stars and
background galaxies. These residuals are removed in the third step. At this stage, they consider 
the data as a mixture of points drawn mainly from two populations, namely the GC and the contaminants.
Thus, the data can be modeled using a mixture model with two component, in which the total 
observed distribution in the $r_h$-$z$  plane is the result
of summing these two components weighted by their respective sizes. The method follows the 
\citet{Fraley:2002} procedure and it is also described in detail in \citet{Jordan:2008}.

\section{Radial Dependencies}\label{radial}

\subsection{Density Profiles of Globular Clusters}\label{profiles}

The study of GCs around giant galaxies within clusters of galaxies has
proved to be an useful tool for improving our knowledge about the
formation and evolution of galaxies and their environments (\citealt{
Hanes:2001}; \citealt{Cote:2001}, \citeyear{Cote:2003}).

The radial distributions of GCs are often fitted with a power law, which can be written as $\log(\rho) =  \alpha\log(r)+\beta$ (\citealt{Harris:1986}, \citealt{AZ:1998}). 
Typically, power law indices range from $\sim -2$ to $\sim -2.5$ for low-luminosity Es \citep{Harris:2001}, and are $\sim -1.5$ for the most massive gEs. In general
terms, blue metal-poor GCs present a shallower density profile than red
GCs. \citet{Bassino:2006} found that both density profiles, for blue
and red GCs, can be well fitted using $\log(\rho)=ar^{1/4}+b$.

We have fitted a power law as well as $r^{1/4}$ law to our galaxy sample. GCs with 
$g-z>1.15 $ were considered metal rich (red GC), while GCs with $g-z<1.15$ were 
considered metal poor (blue GC). The density profiles were computed using equal size bins
($13.5\,\arcsec$), therefore, the number of bins adopted in each galaxy
depends on how far from the galaxy center we detected the GCs.
In some cases only 3 bins were used, while in
others up to 10 bins were considered. Error bars were estimated 
using the bootstrap resampling technique. In Table \ref{tb:fit} the fitting parameters for all the galaxies are quoted. 
The mean value obtained to the slope $\alpha$ for all GCs is $\bar{\alpha}=-2.13\pm 0.08$, while for red and blue GCs are $\bar{\alpha}_{red}=-2.4\pm 0.2$ and {$\bar{\alpha}_{blue}=-1.7\pm 0.1$, respectively. 
Particularly, Fig. \ref{fig:profiles} shows the projected GC density profiles ($r^{1/4}$ law) for 4 galaxies of our sample. 
Left panels show the total GC density profiles, while right panels show the density profiles for the
blue (squares) and red (triangles) GC populations separately. 
VCC1297 shows a flat density profile, therefore no profile was fitted, while 
VCC1422 does not present a red GC population.
For the whole sample, we obtain a mean GC density profile with $\bar{a}=-1.59\pm 0.06$, 
while for the red and blue GC populations the mean values are $\bar{a}_{red}=-1.8\pm 0.1$, 
and $\bar{a}_{blue}=-1.22\pm 0.07$, respectively.
Several galaxies in our sample do not present bimodality in the color 
distribution of GCs (see \citealt{Peng:2006}). Nevertheless, for most galaxies 
in our sample, red GC density profiles are steeper and more centrally
concentrated than the corresponding blue GC density profiles. This could
indicate that we indeed have two different GC populations.

Towards the inner region of the galaxy cluster, tidal effects are expected to
become stronger due to a higher probability of encounters among galaxies, as well
as the increase of the potential well. If we consider, in addition, that GCs located in the outskirts of galaxies are more prone to be stripped, then a correlation between the slope of the
GC density profiles and the distance of hosts galaxies to M87
should be observed. This effect has already been reported by B2003 using
numerical simulations. Nevertheless, for both, blue and red GCs, we do not find any correlation between these 
two parameters (see Fig. \ref{fig:slope}). A possible interpretation to this result is given in Sec. \ref{conclusion}.

\subsection{Specific Frequency $S_N$ and $T$ parameter}

\citet{Harris:1981} introduced the specific frequency $S_N$
as a measure of GC richness normalized to the host galaxy luminosity:
\begin{equation}
S_N=N_{GC}10^{0.4(M_V+15)}
\end{equation}
where $N_{GC}$ is the total number of GCs and $M_V$ is the $V$-band absolute magnitude.
Typical values of $S_N$ are $\sim$1 for spirals and S0 galaxies
\citep{Barmby:2003}, $\sim$3 for ellipticals, and $\sim$6 for cD galaxies \citep{Ostrov:1998}. Nevertheless, 
the correlation between morphological types and $S_N$ contains some dispersion. On the other hand, 
several early studies suggest that $S_N$ increases with galaxy luminosity 
(\citealt{Miller:1998}, \citealt{Lotz:2001}, \citealt{Miller:2007}). However, P2008 found
that the GC mass fraction is high in both giants and dwarfs galaxies,
being universal in intermediate luminosity galaxies, while
\citet{Strader:2006} only found a weak signal in their $S_{N,B}-M_B$ data. 

$S_N$ can also depend on the environment, \citet{F&S:1989} found that the number density of 
nucleated dwarf elliptical galaxies (dE) is more centrally concentrated than that of bright 
non-nucleated dE galaxies in the Virgo and Fornax Clusters. \citet{Miller:2007} did not find a clear 
dependence of $S_N$ on the environment for dEs. Using the VCS, P2008 found that galaxies within 
a projected radius of $1\Mpc$ from M87 tend to have higher GC fractions, and this effect was interpreted 
in terms of the galaxy formation time. It should be noted that the described effect contradicts 
the tidal stripping scenario. Nevertheless, this result is observed when using the total
sample of galaxies and does not discriminate by their luminosity, morphology or location
Virgo's sub-structures.

Panels (a) and (b) of Fig. \ref{fig:Sn} show $S_N$ as a function of ($r_p$) and ($r_{3D}$)
respectively for our sample of elliptical galaxies (see Table \ref{tb:data}).
We observe that larger $S_N$ values are found at larger galaxy distances to M87 ($\sim$ Virgo center). 
This result is indicating that the normalized number of GCs decreases as the host galaxy is closer to the Virgo center, where the tidal stripping is expected to be more efficient. A similar behavior 
is observed in panels (c) and (d), where we plot the specific frequency in the $z$ bandpass $S_{N,z}$, as a 
function of the galaxy distances to M87. In the aforementioned Figure, panels (e),(f) and (g),(h) show the specific frequency $S_{N,z,blue}$ and $S_{N,z,red}$ versus galaxy distances to M87 for the blue and the red GC populations,
respectively. It can be seen that both correlations, projected $S_N$ vs. $r_p$ and $S_N$ vs. $r_{3D}$, are only 
present for the blue GC population. The lack of correlation for the red GCs could be explained by the fact 
that the distribution of red GC is more highly concentrated than the equivalent blue GC distribution, and 
therefore less affected by tidal effects. 
In order to quantify this effect, we split our sample into two:  
galaxies with $r_p<0.8\Mpc$ and $r_p>0.8\Mpc$. The adopted threshold for $r_{3D}$ is $1.3\Mpc$ (assuming that
$r_{3D}=\sqrt{3}r_p$). For each sub-sample, the mean $S_N$ are shown in Table \ref{tb:med}. As it can be seen, galaxies in the outskirts of the Virgo cluster have
mean $S_N$ typically larger than those calculated for galaxies closer to M87. 
Particularly, we excluded VCC 1475 to compute the mean $S_{N,z,blue}$ and the mean $T_{blue}$ due to its unexpected high values.
Alternatively, for the blue GCs we fitted a linear regression between $S_{N,z,blue}$ and $r_p$ and $r_{3D}$ (see panels (e) and (f) in Fig. \ref{fig:Sn}). Table \ref{tb:linearfit} shows the fitting parameters, where it can appreciated that fitted slopes differ from zero at six sigma levels. For the $r_{3D}$ fits, we also excluded VCC 1475.

In addition, we show in Fig. \ref{fig:Sn}, $S_N$ vs. $r_p$ (left panels) and $S_N$ vs. $r_{3D}$ (right panels), discriminating the galaxy sample according to the absolute magnitude $M_z$ of the galaxies. It can be appreciated that the correlation between $S_N$ and $r_p$ and $r_{3D}$ is independent of the galaxy luminosity.

Using stellar masses of galaxies in the VCS, P2008 calculated the $T$ parameter 
introduced by \citet{Z&A:1993}, defined as the number of GCs per $10^9\Msun$,
\begin{equation}
 T=\frac{N_{GC}}{\mathcal{M}_G/10^{9}\Msun}
\end{equation}
where $\mathcal{M}_G$ is the stellar mass of the host galaxy. The use of $T$ allows the comparison across galaxies with different mass-to-light-radius. \citet{Miller:2007} and P2008 found that $T$ correlates 
with galaxy luminosity, in the sense that higher $T$ values correspond to brighter 
galaxies. Nevertheless, in the magnitude range of our sample, no correlation between $T$ and the absolute magnitude is observed. Panels (a) and (b) of Fig. \ref{fig:T} show $T$ vs. $r_p$ and $T$ vs. $r_{3D}$, respectively. 
Panels (c) and (d) correspond to the blue GC population for which we fitted a linear regression 
(again, the $r_{3D}$ fit excludes  VCC 1475), whereas panels (e) and (f) show the corresponding correlations 
for the red GCs. As it was observed for $S_N$, we only find correlations for the blue and the total GC populations.
Higher values of $T$ are found in the outskirts of the Virgo cluster (see also Tab. \ref{tb:med} and Tab. \ref{tb:linearfit}).

\subsection{Local background of GCs}\label{back}

The observed population of GCs is superposed on a background contamination that
exponentially grows with the apparent magnitude limit. This
contamination is basically
constituted by  high redshift galaxies, foreground stars and a projected ICGC
population. If tidal stripping of GCs is an efficient
mechanism in clusters of galaxies, a fraction of the ICGC background
would also be formed by stripped GCs. Moreover, this ICGC background would
show a gradient, having higher values towards smaller projected
clustocentric distances. There are several works that suggest the
existence of the ICGC. \citet{West:1995} suggested that the variation
in the GC population in supergiant elliptical galaxies in clusters
points towards the existence of a GC population that is not bound to
any galaxy and moves throughout the core of the galaxy cluster. This was
also suggested by \citet{Jordan:2003} who used HST images of the rich cluster
A1185. This galaxy cluster contains no bright
galaxies, and shows an excess of point sources that is consistent with
the existence of an ICGC population. The efficiency of GC tidal
stripping due to the cluster potential well can be evaluated in terms
of the tidal radii. Given a galaxy with mass $\mathcal{M}_{gal}$
and at a given distance $r_c$ from the cluster center, the tidal
radius can be computed as
\begin{equation}
r_{tidal}=[G \mathcal{M}_{gal}/2 v_{circ}^2]^{1/3}r_c^{2/3}
\end{equation}
where $v_{circ}$ is the circular velocity of the galaxy cluster
potential. Therefore, for a typical elliptical galaxy with a total mass of
$10^{12}\Msun$ (\citealt{Schindler:1999}), located in the VC potential,
we obtain tidal radii of 11, 14 and 17 $\kpc$ for clustocentric
distances of 0.2, 0.5
and 1.0 $\Mpc$, respectively. These numbers indicate that if a galaxy
passes close
to the VC center it could lose part of its GC population due to tidal stripping.

Numerical simulations run by B2003 and BK2006 suggest that ICGCs
are not necessarily uniformly distributed. GCs stripped from galaxies
have not had enough time to be uniformly distributed in the IC
medium. In fact, the simulation of B2003 shows that a fraction of
the stripped GCs produces an enhancement in the local density around the
parent galaxy. Moreover, BK2006 showed that, on average, ICGCs appear
to have projected radial density profiles that grow towards the galaxy
cluster centers. Based on these two results, we study
the possible dependence of the ICGC population on the galaxy distance to M87.
In our sample, we find that galaxies that are fainter than $M_z=-20.8$
(9 out of 13 galaxies)
have a flat GC projected density profile for galactocentric distances
greater than
$1.3\,\arcmin$ ($6.2 \,\kpc$). These galaxies are suitable to determine the
local GC density
around them since they have relatively small apparent diameters
compared to the ACS field.
We then calculated the local GC density beyond $6.2 \,\kpc$ from the
center of each of these
galaxies. However, this density is still contaminated with high
redshift galaxies and foreground
stars. Hereafter we will refer to this local density as $\rho_{out}$.

P2008 determined fore- and background contamination using 10 blank
fields images
taken with the Wide Field Planetary Camera 2 (WFPC2). They found that
there are on
average $5.8\pm 2.5$ contaminants sources per WFPC2 field. We use this
number to
estimate the contribution of  the mean fore- and background
contamination to $\rho_{out}$.
Figure \ref{fig:icgc} shows the total, blue and red (from top to
bottom) local density
$\rho_{out}$ around the
selected nine galaxies as function of $r_p$ (left panels) and $r_{3D}$
(right panels). Error
bars were estimated using the bootstrap resampling technique.
Horizontal dashed lines
correspond to the mean value of $\rho_{out}$ computed for those
galaxies with clustocentric distances
$r_p>100\,\kpc$. These values are $2.4\pm 0.3$
($\arcmin^{-2}$) for all GCs, $1.7\pm 0.3$ for blue GCs, and $0.6\pm 0.2$
for red GCs. Horizontal dotted lines show the mean density (blue and
red) of the fore- and
background contaminants derived by P2008 from 10 blank fields ($1.1\pm 0.5$).
It can be seen that nine (eight) galaxies
have values of $\rho_{out}$ for the total (blue) GC population
consistent or higher than expected mean values of the fore- and background contaminants. Except for
two galaxies,
the values of  $\rho_{out}$ for the red GC population, are lower than
the expected for the
total mean fore- and background contamination. On the other hand, we
found for the total
and blue GC populations, that three galaxies (VCC1297, VCC1279 and VCC0828) have
$\rho_{out}$ values three times larger than the contaminants mean density.
The same figure also shows that the two highest values of $\rho_{out}$
correspond to
the galaxies with the lowest values of $S_N$ (VCC1297 and VCC1279) that
are also the galaxies closest to M87 (see Table \ref{tb:data}), which has an extended halo
of GCs. The
extended GC halo of M87 could be partially contributing to the high
density levels observed
around these two galaxies.
However, since M87 is located near the center of the cluster potential
well we are unable to
determine if this additional contribution is derived from GCs bound to
M87 itself, or if they are
truly ICGCs. In any case, the procedure we use to compute the
background is independent
of the origin of GCs and therefore, our method does accurately
represent the GC background
around these galaxies.

\section{Conclusions and Discussion}\label{conclusion}

We analysed a sample of 13 elliptical galaxies taken from the ACS VCS
with the aim of finding evidence of the tidal stripping of GCs. 
Particularly, we focused our attention on the dependence of several galaxy properties
with the radial clustocentric distance. The selected sample comprises host galaxies in the magnitude 
range $-22.5<M_z<-18.5$ that belong to the main concentration of the VC. The GCs around these galaxies were taken from \citet{Jordan:2008}, while several galaxy properties were taken from P2008. The color index $g-z$ was used to discriminate metal poor (blue) from metal rich (red) GCs.

There is a correlation between the specific frequency, in the $V$ and $z$-bands,
and the galaxy distances to M87. Larger values of $S_{N}$ are found at larger clustocentric distances. 
The effect is observed using both projected or 3-dimensional distances. If GCs are selected according to their color, 
the correlation is only observed in the blue (metal poor) population. These results are in agreement 
with the idea that those galaxies closer to the Virgo center are loosing a fraction of their GCs 
due to tidal stripping. The lack of correlation for the red GC (metal rich) 
population could be explained if we consider that these objects have a more concentrated 
distribution around the host galaxy, and therefore they are less affected by tidal effects than 
the blue GCs. Similarly, we found a correlation between the $T$ parameter (the number of GCs normalized to galaxy stellar mass) and $r_p$ and $T$ and $r_{3D}$, for the total and blue GC populations only. $T$ increases as the galaxy distance to M87 increases. The lowest values of $S_N$ and $T$ are found within $100\kpc$ 
to the cluster center, where P2008 have also reported evidences of tidal stripping. 

As it was explained before, P2008 found that: a) $S_N$ and $T$ depend on the
environment, i.e. galaxies within a projected radius of 1 Mpc from the
cluster center tend to have higher GC fractions than galaxies at larger clustocentric distances. 
b) Galaxies within $r_p\simeq 40\,\kpc$ from M87 (and M49) have a few or no GCs.
The first result is interpreted in terms of formation time. GC formation in
centrally located low-mass galaxies is biased because their stars form earlier. The second result 
considers that GCs are tidally stripped from the parent galaxies by their giant neighbors.
At first glance, our findings appear to differ from the aforementioned P2008 result (a),
since we find that the mean $S_N$ value is indeed higher at larger $r_p$. 
Nevertheless, before we can compare our results to those of P2008 we have to consider
the following: i) although all VCC galaxies are early-types, P2008 analysed the radial 
dependence for a great variety of galaxy morphologies, while our analyses only consider pure elliptical galaxies;
ii) our sample excludes bright and faint VCC galaxies and comprises only those objects in the magnitude 
range $-22.5<M_z<-18.5$, for which P2008 data show no correlation between the galaxy magnitude and $S_N$. 
This is the most important difference between our analysis and that performed by P2008, since 
the dependences of $S_N$ and $T$ on the galaxy magnitude tend to erase any other correlation;
iii) our study only considers galaxies that belong to the main concentration around M87 and excludes 
galaxies close to M49. It is more likely that the dynamic history of these galaxies is more 
related to the group of M49 than to the Virgo cluster.

We tested a possible correlation between the slopes of the GC density profiles and the 
clustocentric distances as predicted by B2003 numerical simulations. Computed slopes
for our galaxy sample do not show correlation with $r_p$ nor with $r_{3D}$. 
\citet{Cypriano:2006} found that galaxies in the inner regions of clusters
are 5\% smaller than those in the outer regions. These authors interpret this result in terms 
of star stripping. \citet{A&W:1986,A&W:1987} studied this effect using numerical simulations 
and they found that the $r^{1/4}$ luminosity profile is robust, since it is recovered after
the galaxy interaction or merger. However,  $r^{1/4}$ paramenters, the effective radius $r_{e}$ and the surface brightness $\mu_{e}$ inside $r_{e}$, change
depending on the collision type. Strong collisions result in a reduced $r_{e}$
and a higher $\mu_{e}$, while weak collisions produce the opposite effect.
If GCs are affected by the same process that affect the stars, the GC distribution 
could also be modified, but the final density profile will also depend on the \index{}interaction class.
If this model is correct, no correlation between the slopes of the GC distribution and the clustocentric distances
should be expected. 

We also computed the GC density $\rho_{out}$, outside $6.2 \,\kpc$
from the galaxy center for
nine galaxies of our sample. Our results show that values of
$\rho_{out}$ for the total (blue)
GC population around nine (eight) of the selected fields are
consistent, or larger, than those
expected from the fore and background contaminants. Except for two of
the nine selected fields,
the values of $\rho_{out}$ for the red GC population are lower than expected for the mean
fore- and background contamination.
These results indicate that the ICGC population around most of the
galaxies of our sample is
mainly composed of blue GC.
The two highest values of $\rho_{out}$ (that correspond to the two
lowest values of $S_N$)
are found in the core of the VC (within $100\kpc$),
confirming P2008 results.
They could be stripped GCs from neighbor galaxies, some fraction of
the very extended GC distribution of
M87, and/or a genuine ICGC population.
The two fields at larger clustocentric distances with total and blue
values of $\rho_{out}$ larger
than expected for the fore- and background contamination could be
either a cosmic variance
or the result of GCs stripped from the parent galaxy that have not had
enough time to uniformly distribute
into the intracluster medium.

The evidence for tidal stripping found in this work suggests that special care should be taken
in studies that attempt to model the formation and evolution of the GC population. 
The number, proportion and distribution of GCs observed in those elliptical galaxies close to the center 
in a massive cluster, appear to significantly differ from the GC population in galaxies before
they approached the center of the cluster. 
Moreover, a good estimate of the background contamination in GC counts is critical, since 
we have shown the possible existence of both, a mean gradient as a function of $r_p$, 
and a possible cosmic scatter of the mean local background around some galaxies. 
Finally, more realistic simulations are needed in order to confront observational results with
model predictions. A complete set of cluster galaxies selected from a cosmological simulation 
should be analyzed. Besides, metal poor and metal rich GCs with their corresponding radial 
distributions should be considered. Simulations constrained in this way, have the potential to
predict the initial undisturbed radial distribution of GCs around galaxies in rich clusters.

\section{Acknowledgments}
We kindly thank Sebastian Gurovich and Eugenia D\'iaz for their helpful comments on the manuscript.
We are grateful to the referee for his valuable comments, which contribute
to improve the present paper. HM thanks Juan Carlos
Forte and Favio Faifer for the useful discussions.  This work was
partially supported by the Consejo de Investigaciones Cient\'{\i}ficas
y T\'ecnicas de la Rep\'ublica Argentina, CONICET; SeCyT, UNC, Agencia
Nacional de Promoci\'on Cient\'{\i}fica, Argentina. Support for this
work was provided by the National Science Foundation through grant
N1183 from the Association of Universities for Research in Astronomy,
Inc., under NSF cooperative agreement AST-0132798.\\

\acknowledgments

\bibliographystyle{apj} 
\bibliography{ms} 

\clearpage



\begin{figure}
\center
	\subfigure{\includegraphics[width=8cm]{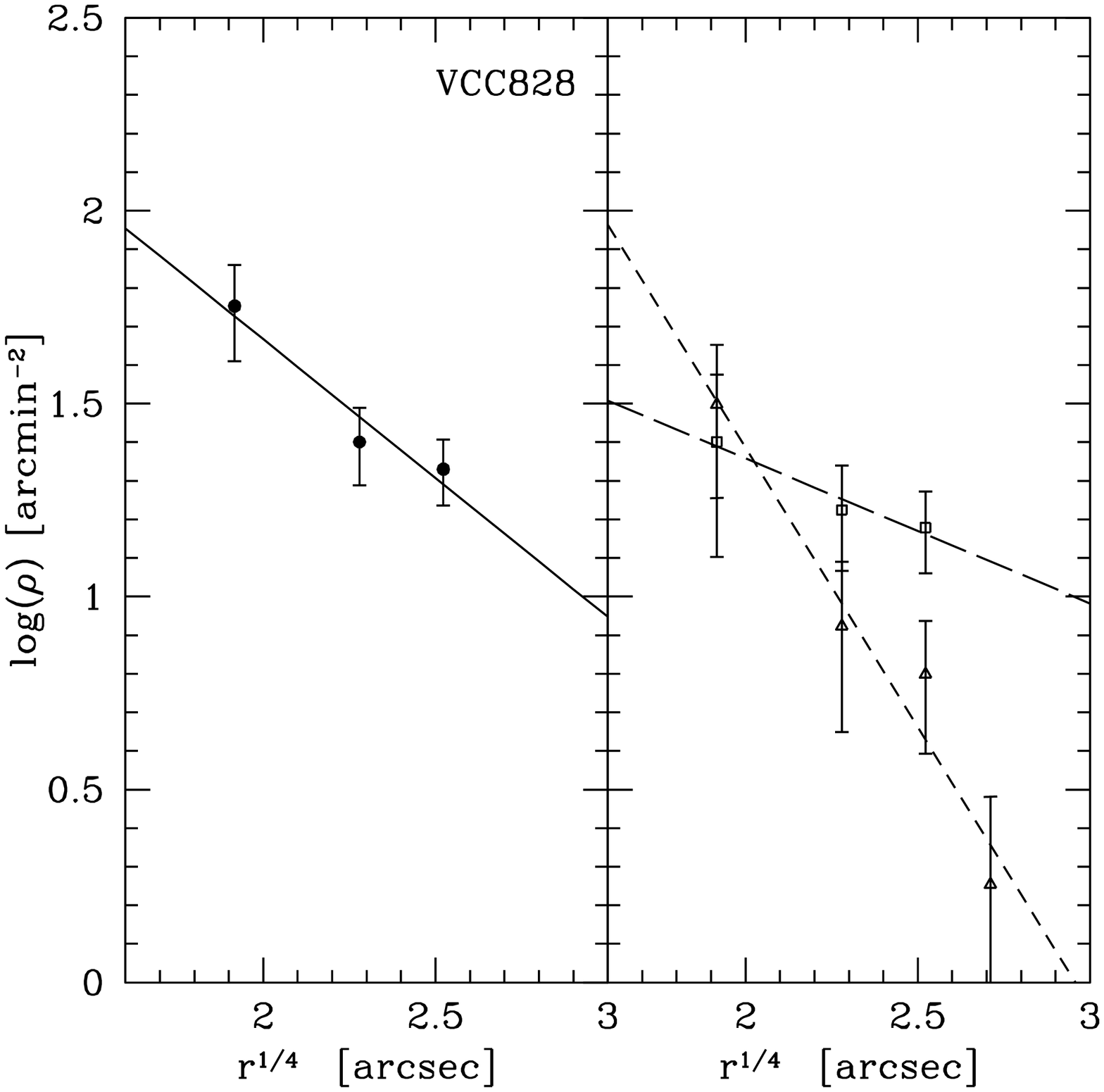}}
	\subfigure{\includegraphics[width=8cm]{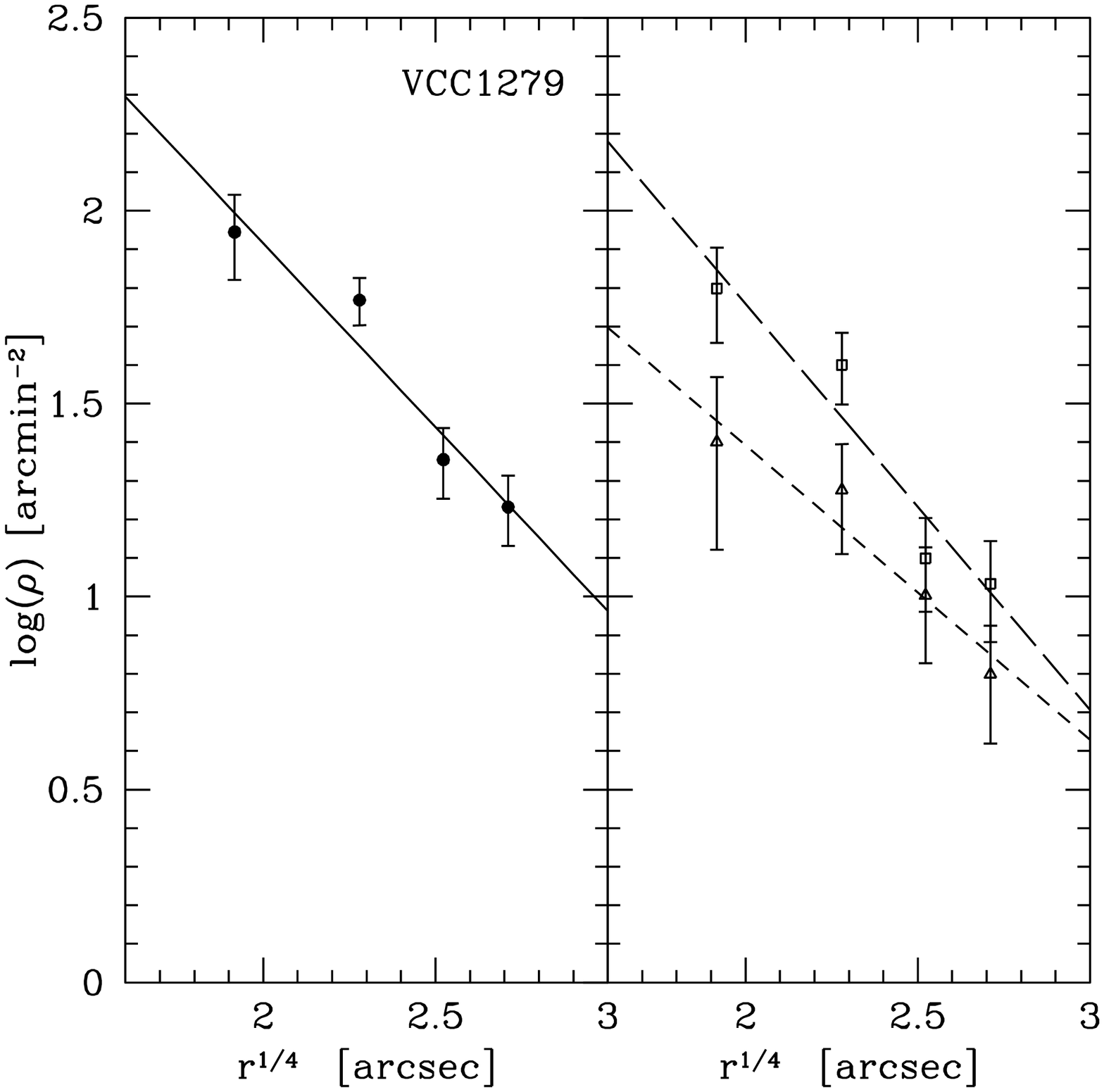}}\\
	\subfigure{\includegraphics[width=8cm]{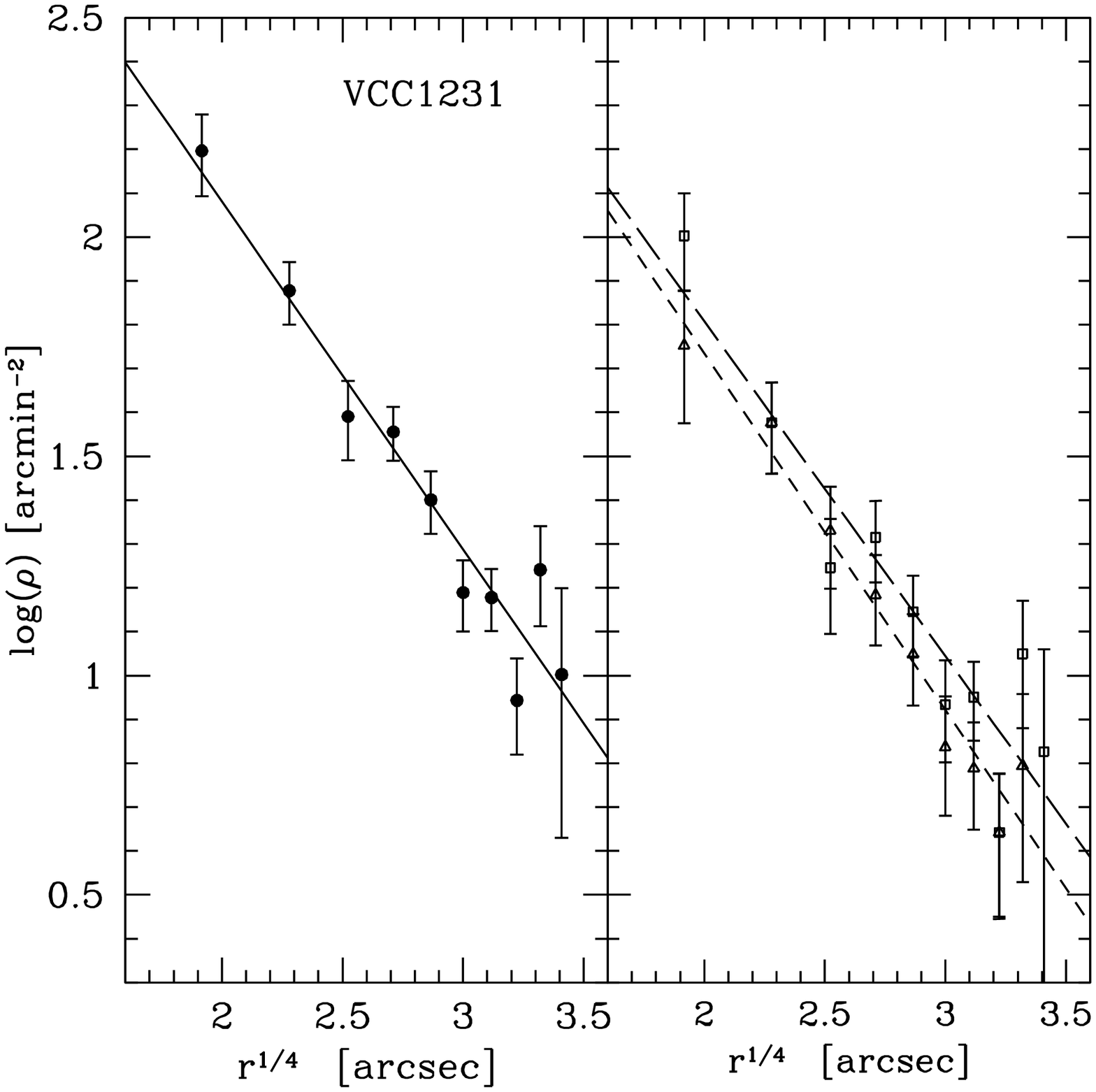}}
	\subfigure{\includegraphics[width=8cm]{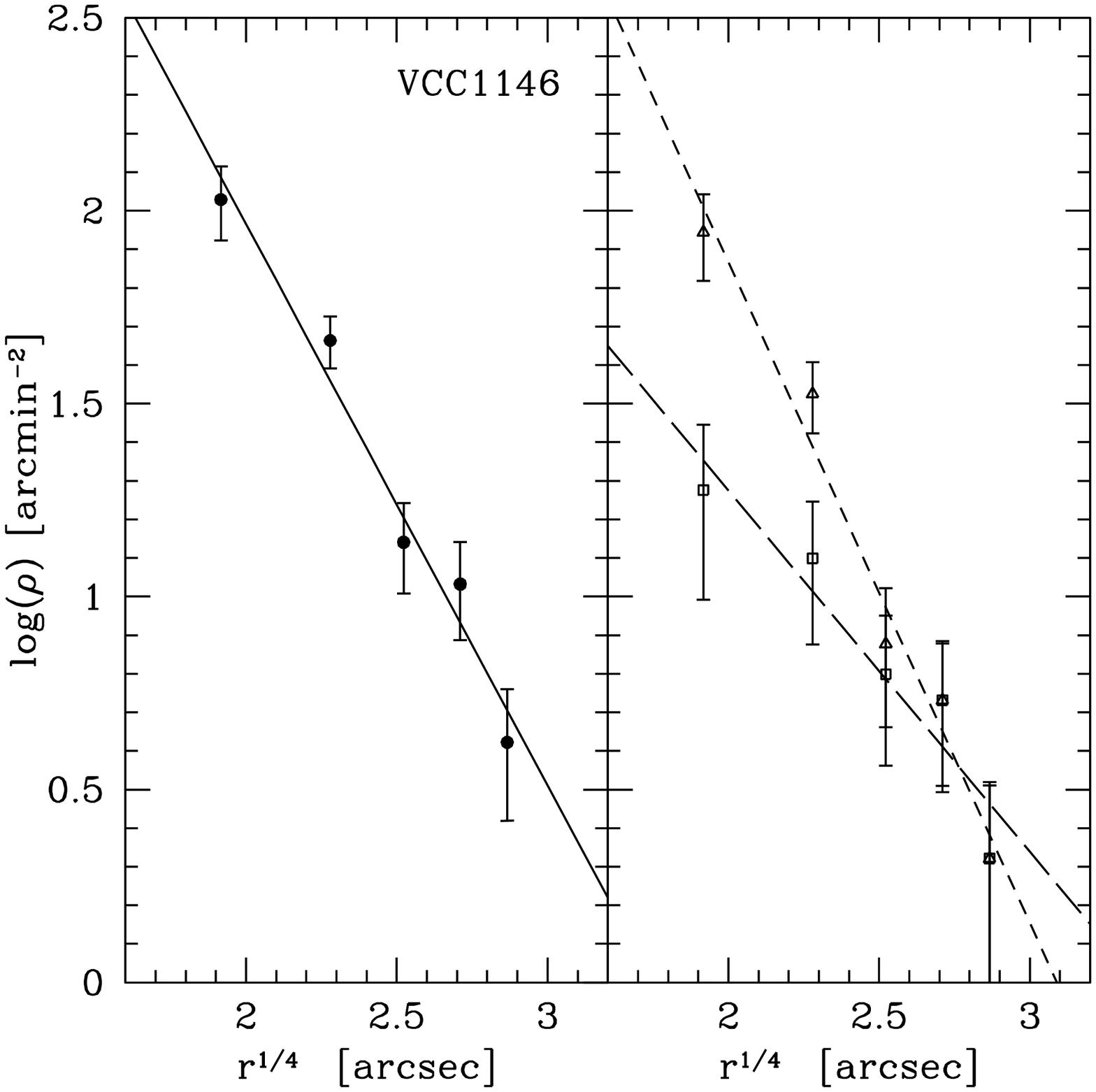}}
	\caption{GC projected density profiles for four galaxies in our sample. Filled 
squares, open triangles and open squares correspond to the total, red and blue GC, 
respectively. \emph{Long dashed lines} and \emph{short dashed lines} correspond to $r^{1/4}$ fit for the blue and the red GC population, respectively. Errors bars were estimated using the bootstrap resampling technique.}
	\label{fig:profiles}
\end{figure}

\begin{figure}
\includegraphics[scale=.85]{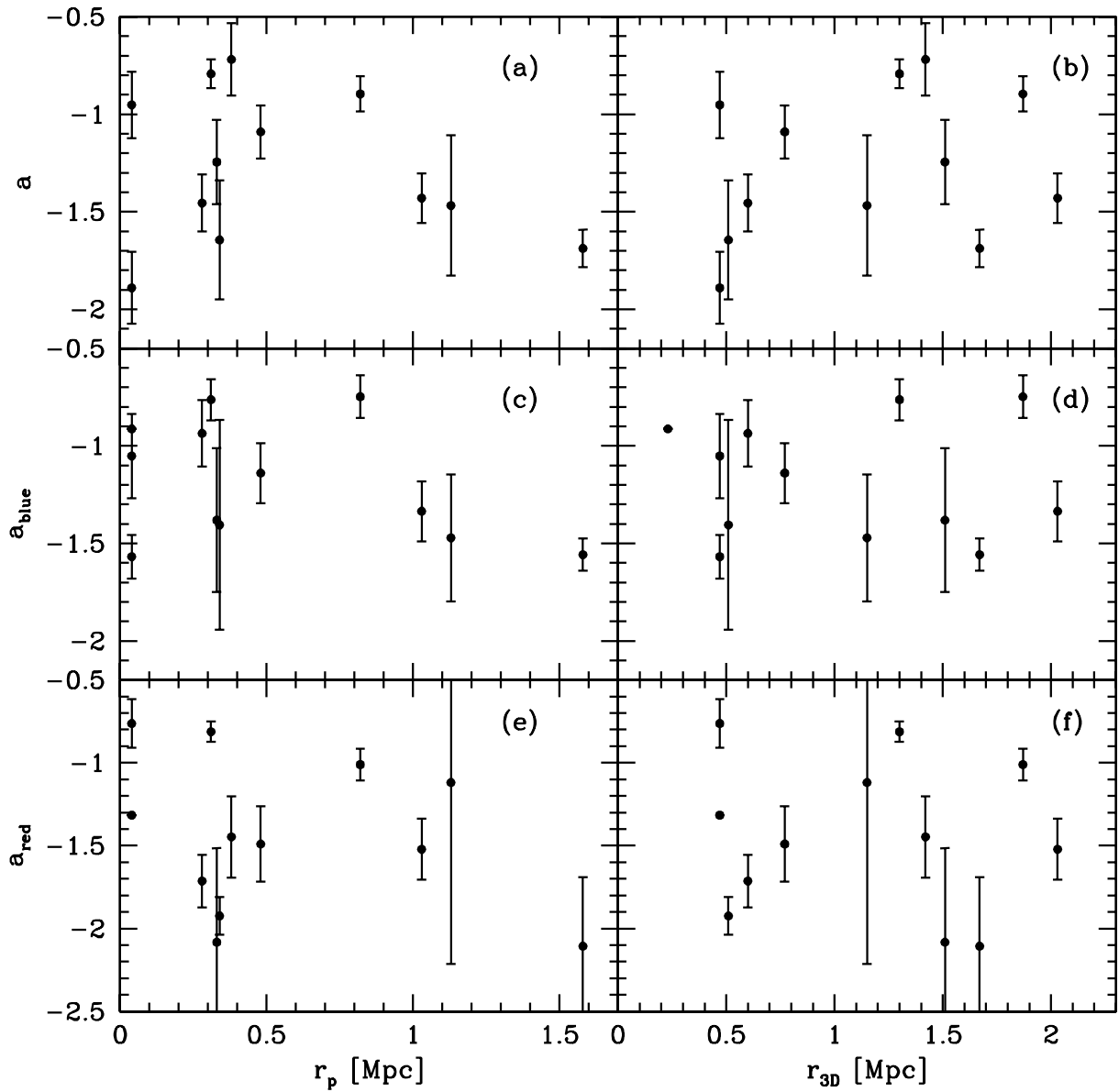}
\caption{Slopes of the GC projected density profiles as a function of the galaxy projected distance $r_p$ (\emph{left panels}) and as a funtion of the galaxy 3-dimensional distance $r_{3D}$ to M87 (\emph{right panels}). \emph{From top to bottom}: correlations for the total GC population, blue and red GC pupulations.}
\label{fig:slope}
\end{figure}

\begin{figure}
\includegraphics[scale=.85]{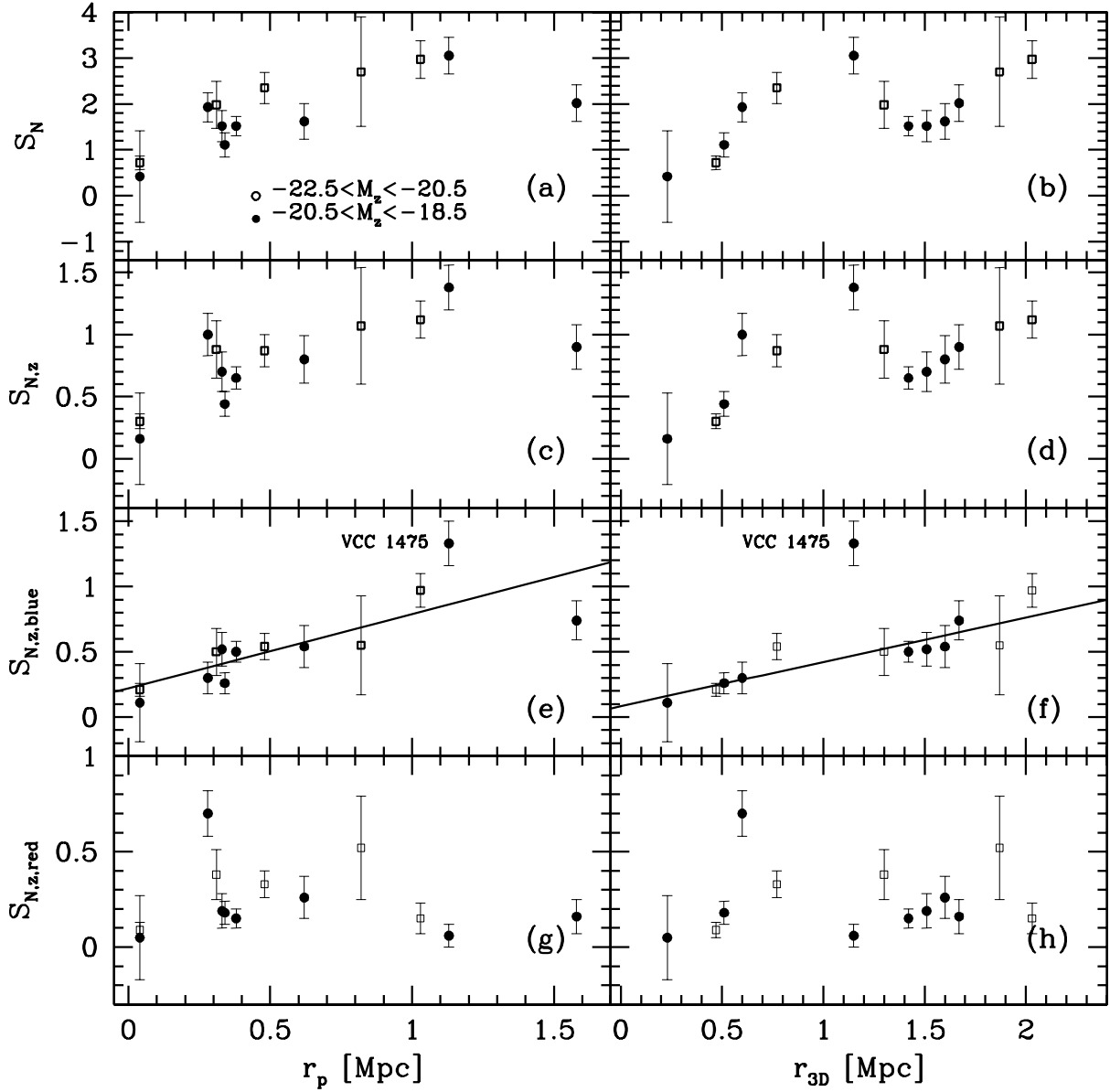}
\caption{$S_N$ of each galaxy as a function of the clustocentric projected distance $r_p$ of the host galaxy (\emph{left panels}) and as a funtion of the galaxy 3-dimensional distance $r_{3D}$ (\emph{right panels}). \emph{From top to bottom}: correlations for the total GC population in $V$-band, total GC population in the $z$-band, blue GC pupulation in the $z$-band and red GC population in the $z$-band. \emph{Black solid lines} in panels e) and f) correspond to linear regressions. \emph{Open and solid symbols} consider two different ranges in the absolute magnitude $M_z$ of the host galaxy.}
\label{fig:Sn}
\end{figure}

\begin{figure}
\includegraphics[scale=.85]{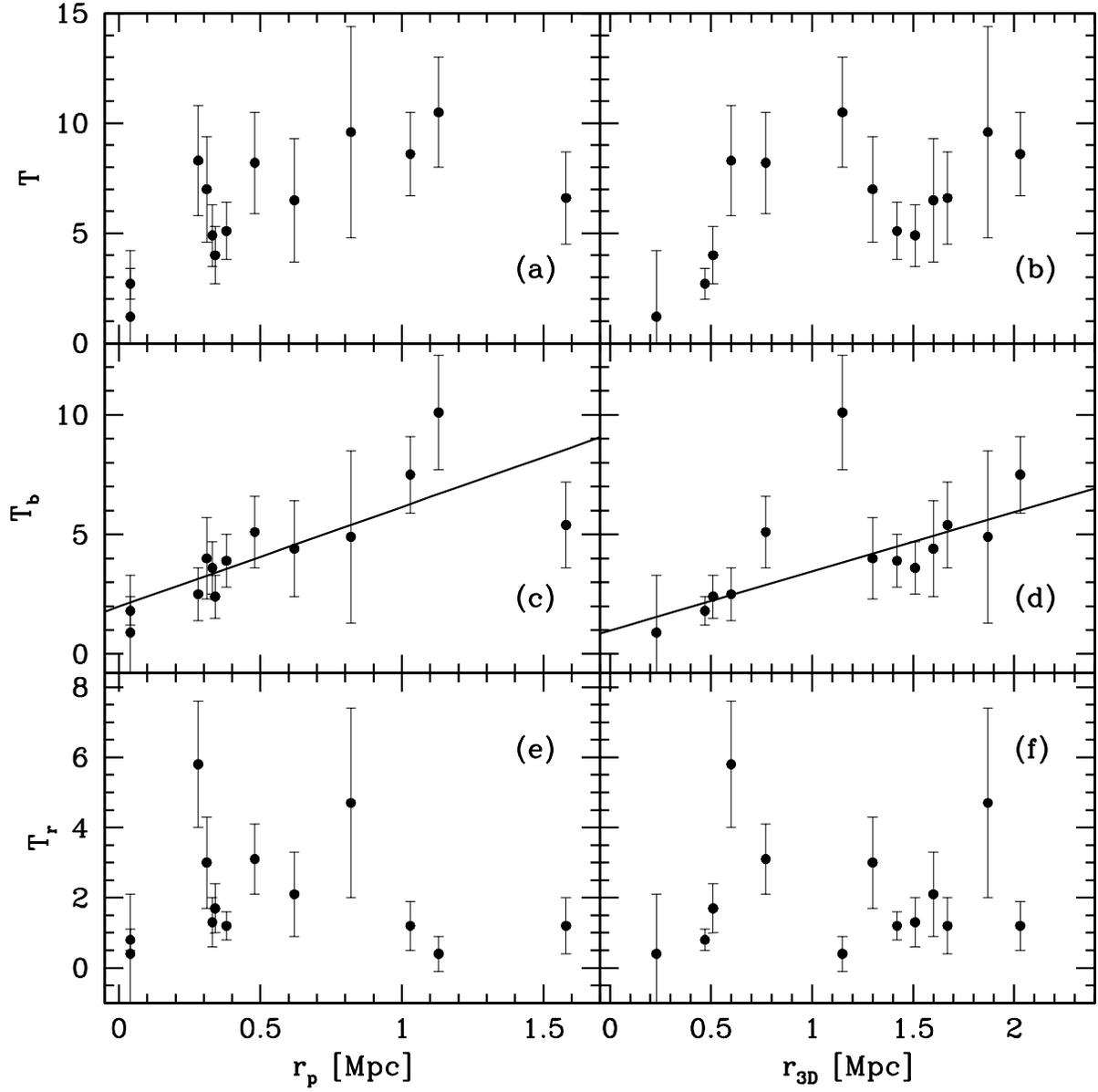}
\caption{Some as Fig. \ref{fig:slope} but now for the $T$ parameter.}
\label{fig:T}
\end{figure}

\begin{figure}
\includegraphics[scale=.85]{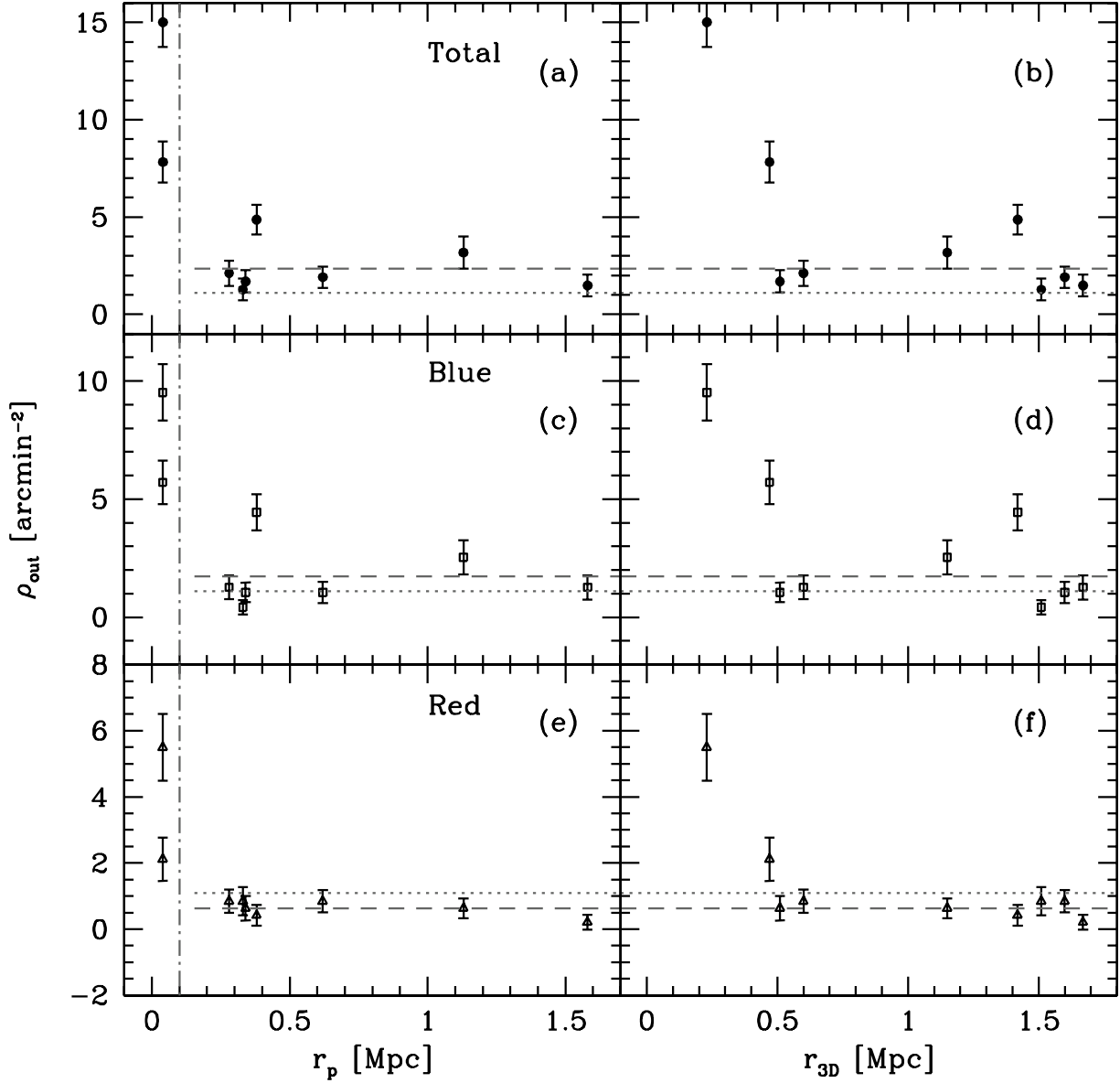}
\caption{Total local density $\rho_{out}$ as a function of the distance to M87, \emph{from top to bottom}: for the total GC population, blue GC pupulation and red GC population. \emph{Black solid lines} in panels e) and f) correspond to linear regressions. The horizontal dashed lines correspond to the mean total density computed around galaxies at $r_p>100\,\kpc$ 
(\emph{vertical lines}). Dotted lines show the expected mean density of the fore- and background contaminants. Errors bars were estimated using the bootstrap resampling technique.}
\label{fig:icgc}
\end{figure}


\begin{deluxetable}{ccccccccccccc}
\tabletypesize{\tiny}
\rotate
\tablecaption{Galaxy properties in ACS Virgo Cluster Survey taken from \citet{Peng:2008} and \citet{Mei:2007}.\label{tb:data}}
\tablewidth{0pc}
\tablehead{
\colhead{VCC} &
\colhead{$M_z$} &
\colhead{$r_p$} &
\colhead{$r_{3D}$} &
\colhead{$N_{GC}$} &
\colhead{$S_N$} &
\colhead{$S_{N,z}$} &
\colhead{$S_{N,z,blue}$} &
\colhead{$S_{N,z,red}$} &
\colhead{$T$} &
\colhead{$T_{blue}$} &
\colhead{$T_{red}$} & 
\colhead{Type} \\
\colhead{(1)} &
\colhead{(2)} &
\colhead{(3)} &
\colhead{(4)} &
\colhead{(5)} &
\colhead{(6)} &
\colhead{(7)} &
\colhead{(8)} &
\colhead{(9)} &
\colhead{(10)} &
\colhead{(11)} &
\colhead{(12)} &
\colhead{(13)} \\
}
\startdata
1903 &    $-22.19$ & $0.82$ & $1.87\pm0.35$ & $803\pm355$ & $2.71\pm1.19$ & $1.07\pm0.47$ & $0.55\pm0.38$ & $0.52\pm0.27$ & $9.6\pm4.8$ & $6.2\pm2.0$ & $4.1\pm1.4$ & E5 \\ 
1231 &    $-21.58$ & $0.31$ & $1.30\pm0.36$ & $376\pm97$ & $1.98\pm0.51$ & $0.88\pm0.23$ & $0.50\pm0.18$ & $0.38\pm0.13$ & $7.0\pm2.4$ & $4.0\pm1.7$ & $3.0\pm1.3$ & E5\\ 
2000 &    $-20.66$ & $1.03$ & $2.03\pm0.35$ & $205\pm28$ & $2.97\pm0.41$ & $1.12\pm0.15$ & $0.97\pm0.13$ & $0.15\pm0.08$ & $8.6\pm1.9$ & $7.5\pm1.6$ & $1.2\pm0.7$ & E5\\ 
1664 &    $-20.97$ & $0.48$ & $0.77\pm0.38$ & $213\pm31$ & $2.35\pm0.34$ & $0.87\pm0.13$ & $0.54\pm0.10$ & $0.33\pm0.07$ & $8.2\pm2.3$ & $5.1\pm1.5$ & $3.1\pm1.0$ & E6\\ 
1279 &    $-20.72$ & $0.04$ & $0.47\pm0.40$ & $58\pm11$ & $0.72\pm0.15$ & $0.30\pm0.06$ & $0.21\pm0.15$ & $0.09\pm0.04$ & $2.7\pm0.7$ & $1.8\pm0.6$ & $0.8\pm0.3$ & E2\\ 
1619 &    $-20.19$ & $0.33$ & $1.51\pm0.36$ & $84\pm19$ & $1.52\pm0.34$ & $0.70\pm0.16$ & $0.52\pm0.13$ & $0.19\pm0.09$ & $4.9\pm1.4$ & $3.6\pm1.1$ & $1.3\pm0.7$ & E/S0\\ 
0828 &    $-20.07$ & $0.38$ & $1.42\pm0.42$ & $69.5\pm9.8$ & $1.52\pm0.21$ & $0.65\pm0.09$ & $0.50\pm0.08$ & $0.50\pm0.08$ & $5.1\pm1.3$ & $3.9\pm1.1$ & $1.2\pm0.4$ & E5\\ 
1630 &    $-22.19$ & $0.82$ & $1.87\pm0.35$ & $47\pm11$ & $1.11\pm0.26$ & $0.44\pm0.10$ & $0.26\pm0.08$ & $0.18\pm0.06$ & $4.0\pm1.3$ & $2.4\pm0.9$ & $1.7\pm0.7$ & E\\ 
1146 &    $-20.06$ & $0.34$ & $0.51\pm0.38$ & $72\pm12$ & $1.93\pm0.32$ & $1.00\pm0.17$ & $0.30\pm0.12$ & $0.70\pm0.12$ & $8.3\pm2.5$ & $2.5\pm1.1$ & $5.8\pm1.8$ & E\\ 
1913 &    $-19.74$ & $1.58$ & $1.67\pm0.40$ & $71\pm14$ & $2.02\pm0.40$ & $0.90\pm0.18$ & $0.74\pm0.15$ & $0.16\pm0.09$ & $6.6\pm2.1$ & $5.4\pm1.8$ & $1.2\pm0.8$ & E\\ 
1475 &    $-19.42$ & $1.13$ & $1.15\pm0.38$ & $81\pm10$ & $3.05\pm0.40$ & $1.38\pm0.18$ & $1.33\pm0.17$ & $0.06\pm0.06$ & $10.5\pm2.5$ & $10.1\pm2.4$ & $0.4\pm0.5$ & E\\ 
1422 &    $-18.73$ & $0.62$ & $1.60\pm0.43$ & $24.9\pm6.0$ & $1.62\pm0.39$ & $0.80\pm0.19$ & $0.54\pm0.16$ & $0.26\pm0.11$ & $6.5\pm2.8$ & $4.4\pm2.0$ & $2.1\pm1.2$ & E\\ 
1297 &    $-18.75$ & $0.04$ & $0.23\pm0.46$ & $4\pm11$ & $0.42\pm1.00$ & $0.16\pm0.37$ & $0.11\pm0.30$ & $0.05\pm0.22$ & $1.2\pm3.0$ & $0.9\pm2.4$ & $0.4\pm1.7$ & E\\ 
\enddata
\tablecomments{(1) Number in Virgo Cluster Catalog. (2) Absolute $z$ magnitude. (3) Projected distance from M87, in Mpc. (4) 3-dimensional distance from M87, in Mpc. (5) Total number of GCs. (6) Specific frequency in $V$-band. (7) Specific frecuency in $z$ bandpass. (8) Specific frecuency in $z$ for blue GCs. (9) Specific frecuency in $z$ for red GCs. (10) $T$ parameter: $N_{GC}$ normalized to stellar mass of $10^9M_{\odot}$. (11) $T$ parameter to blue GCs. (12) $T$ parameter to red GCs. (13) Morphological type.}
\end{deluxetable}

\begin{deluxetable}{ccccccc}
\tabletypesize{\small}
\tablecaption{Fit parameters of density profiles. \label{tb:fit}}
\tablewidth{0pc}
\tablehead{
\colhead{VCC} &
\colhead{$a$} &
\colhead{$a_{blue}$} &
\colhead{$a_{red}$} &
\colhead{$\alpha$} &
\colhead{$\alpha_{blue}$} &
\colhead{$\alpha_{red}$} \\
}
\startdata
1903 &    -0.90$\pm$0.09 & -0.7$\pm$0.1   & -1.0$\pm$0.1   &  -1.3$\pm$0.2 & -1.1$\pm$0.2   & -1.4$\pm$0.2 \\ 
1231 &    -0.79$\pm$0.07 & -0.8$\pm$0.1   & -0.81$\pm$0.06 &  -1.2$\pm$0.1 & -1.2$\pm$0.1   & -1.2$\pm$0.1 \\
2000 &    -1.4$\pm$0.1   & -1.3$\pm$0.2   & -1.5$\pm$0.2   &  -2.1$\pm$0.3 & -2.0$\pm$0.3   & -2.1$\pm$0.3 \\
1664 &    -1.1$\pm$0.1   & -1.1$\pm$0.2   & -1.5$\pm$0.2   &  -1.7$\pm$0.2 & -1.7$\pm$0.2   & -2.1$\pm$0.4 \\
1279 &    -1.0$\pm$0.2   & -1.1$\pm$0.2   & -0.8$\pm$0.1   &  -1.2$\pm$0.3 & -1.4$\pm$0.3   & -1.0$\pm$0.2 \\
1619 &    -102$\pm$0.2   & -1.4$\pm$0.4   & -2.1$\pm$0.7   &  -1.7$\pm$0.3 & -1.9$\pm$0.5   & -2.6$\pm$0.8 \\
0828 &    -0.7$\pm$0.2   & -0.38$\pm$0.08 & -1.4$\pm$0.2   &  -0.9$\pm$0.2 & -0.48$\pm$0.09 & -1.9$\pm$0.4 \\
1630 &    -1.6$\pm$0.3   & -1.4$\pm$0.5   & -1.9$\pm$0.1   &  -2.1$\pm$0.4 & -1.8$\pm$0.7   & -2.5$\pm$0.2 \\
1146 &    -1.5$\pm$0.1   & -0.9$\pm$0.1   & -1.7$\pm$0.2   &  -1.9$\pm$0.2 & -1.2$\pm$0.3   & -2.3$\pm$0.3 \\
1913 &    -1.7$\pm$0.1   & -1.56$\pm$0.08 & -2.1$\pm$0.4   &  -2.4$\pm$0.1 & -2.1$\pm$0.1   & -2.8$\pm$0.5 \\
1475 &    -1.5$\pm$0.4   & -1.5$\pm$0.3   & -1$\pm$1       &  -1.9$\pm$0.5 & -1.9 $\pm$0.5  & -1.4$\pm$1.5  \\
1422 &    -1.9$\pm$0.2   & -1.6$\pm$0.1   &  ...           &  -2.5$\pm$0.3 & -1.99$\pm$0.07 &   ...         \\
1297 &      ...          &     ...        &  ...           &    ...        &    ...         &   ...          \\ 
\hline
\emph{mean values} & -1.59$\pm$0.06   & -1.22$\pm$0.07   & -1.8$\pm$0.1 & -1.13$\pm$0.08 & -1.7 $\pm$0.1  & -2.4$\pm$0.2  \\
\enddata
\end{deluxetable}


\begin{deluxetable}{ccccc}
\tablecaption{Mean values of $S_N$ and $T$ as a function of $r_p$ and $r_{3D}$. \label{tb:med}}
\tablewidth{0pc}
\tablehead{
\colhead{} &
\colhead{$r_p<0.8 \, \Mpc$} &
\colhead{$r_p>0.8 \, \Mpc$} &
\colhead{$r_{3D}<1.3 \, \Mpc$\tablenotemark{a}} &
\colhead{$r_{3D}>1.3 \, \Mpc$} \\
}
\startdata
$S_N$            &    $1.5\pm 0.2$     &  $2.4\pm 0.3$   &    $1.8\pm 0.2$     &    $2.0\pm 0.4$ \\
$S_{N,z}$        &    $0.67\pm 0.07$   &  $1.0\pm 0.1$   &    $0.78\pm 0.06$   &    $0.9\pm 0.2$\\
$S_{N,z,blue}$\tablenotemark{b}   &    $0.40\pm 0.06$   &  $0.77\pm 0.09$ &    $0.45\pm 0.06$   &    $0.7\pm 0.2$\\
$S_{N_,z,red}$   &    $0.3\pm 0.04$    &  $0.2\pm 0.06$  &    $0.28\pm 0.04$   &    $0.2\pm 0.09$\\
$T$              &    $5.5\pm 0.3$     &  $8\pm 1$       &    $6.4\pm 0.7$     &    $7\pm 2$   \\
$T_{blue}$\tablenotemark{b}       &    $3.3\pm 0.5$     &  $6.3\pm 0.8$   &    $3.8\pm 0.5$     &    $5\pm 2$\\
$T_{red}$        &    $2.2\pm 0.3$     &  $1.7\pm 0.4$   &    $2.1\pm 0.3$     &    $1.2\pm 0.8$\\
\enddata
\tablenotetext{a}{\scriptsize $1.3\Mpc$ correspond to $r_{3D}\sim 0.8\sqrt{3} \Mpc$.}
\tablenotetext{b}{\scriptsize In the $r_{3D}$ analysis, we excluded VCC1475 (see the text).}
\end{deluxetable}

\begin{deluxetable}{cccc}
\tabletypesize{\small}
\tablecaption{Parameters of the linear fit of $S_{N,z,blue}$ and $T_{blue}$ as function of $r_p$ and $r_{3D}$. $a$ is the slope and $b$ the zeropoint. \label{tb:linearfit}}
\tablewidth{0pc}
\tablehead{
\colhead{} &
\colhead{} &
\colhead{$r_p$} &
\colhead{$r_{3D}$} \\
}
\startdata
               & $a$ & $0.6\pm 0.1$ & $0.34\pm 0.06$ \\
$S_{N,z,blue}$ &     &              &                \\
               & $b$ & $0.2\pm 0.1$ & $0.08\pm 0.07$ \\
\hline
               & $a$ & $4\pm 1$     & $2.5\pm 0.5$   \\
$T_{blue}$     &     &              &                \\
               & $b$ & $2.0\pm 0.7$ & $1.0 \pm 0.6$  \\
\enddata
\end{deluxetable}

\end{document}